\documentclass[journal]{vgtc}              

\newcommand{\tool}{{\texttt{FactFlow}}\xspace}

\onlineid{1040}

\vgtccategory{Research}

\vgtcpapertype{application/design study}

\title{\tool: Automatic Fact Sheet Generation and Customization from Tabular Dataset via AI Chain Design \& Implementation}

\author{%
  Minh Duc Vu, Jieshan Chen, Zhenchang Xing, Qinghua Lu, Xiwei Xu, and Qian Fu
}

\authorfooter{
  \item
  	Minh Duc Vu, Jieshan Chen, Zhenchang Xing, Qinghua Lu, Xiwei Xu, and Qian Fu are with CSIRO's Data61.
  	E-mail: firstname.lastname@data61.csiro.au
  \item
  	Zhenchang Xing is also with Australian National University.
  \item
        Jieshan Chen is the corresponding author.
}

\abstract{%
With the proliferation of data across various domains, there is a critical demand for tools that enable non-experts to derive meaningful insights without deep data analysis skills. To address this need, existing automatic fact sheet generation tools offer heuristic-based solutions to extract facts and generate stories. However, they inadequately grasp the semantics of data and struggle to generate narratives that fully capture the semantics of the dataset or align the fact sheet with specific user needs. Addressing these shortcomings, this paper introduces \tool, a novel tool designed for the automatic generation and customisation of fact sheets. \tool applies the concept of collaborative AI workers to transform raw tabular dataset into comprehensive, visually compelling fact sheets. We define effective taxonomy to profile AI worker for specialised tasks. Furthermore, \tool empowers users to refine these fact sheets through intuitive natural language commands, ensuring the final outputs align closely with individual preferences and requirements. Our user evaluation with 18 participants confirms that \tool not only surpasses state-of-the-art baselines in automated fact sheet production but also provides a positive user experience during customization tasks.
}


\teaser{
  \centering
  \includegraphics[width=\linewidth]{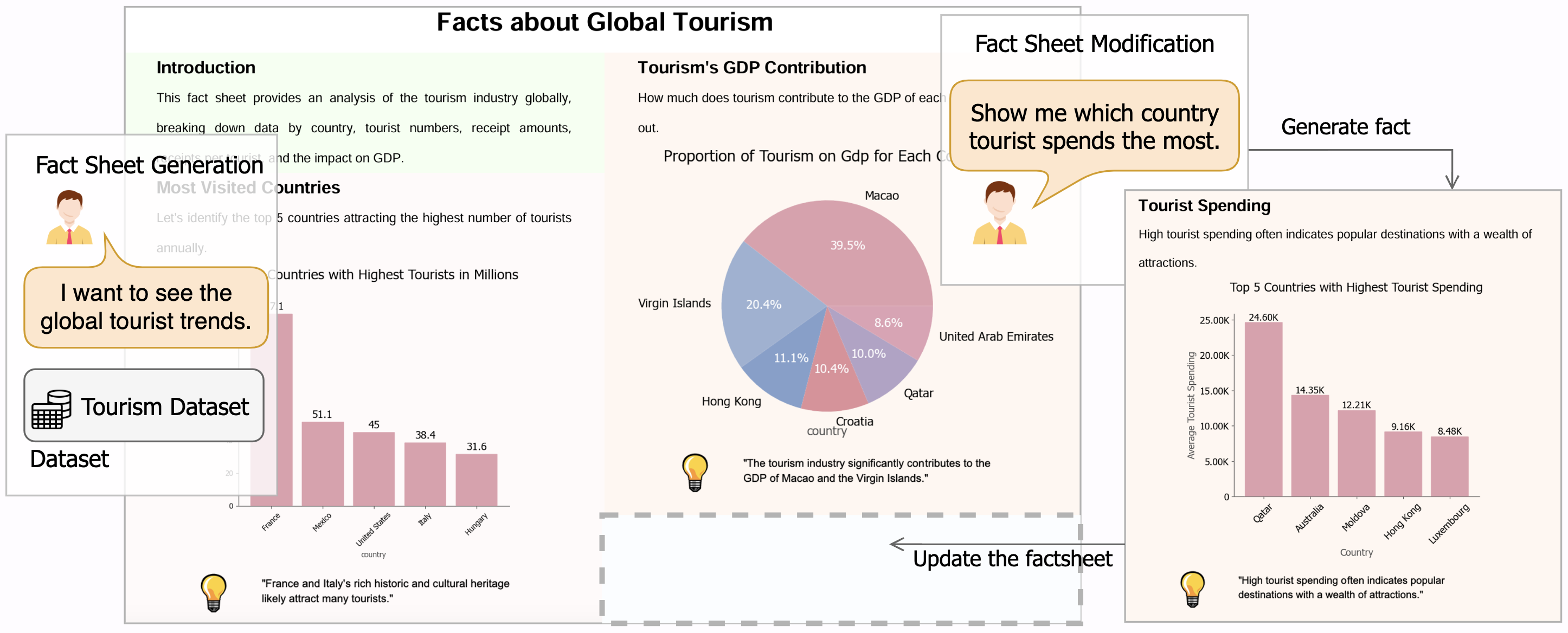}
\caption{%
We demonstrate \tool's functionality using the Tourism dataset. \tool first processes the dataset and the user's initial ideas to automatically generate a fact sheet. Users can then dynamically modify the fact sheet through natural language requests, prompting \tool to generate new content and update the fact sheet accordingly.
}
  \label{fig:teaser}
}

\graphicspath{{figs/}{figures/}{pictures/}{images/}{./}} 

\usepackage{tabu}                      
\usepackage{booktabs}                  
\usepackage{lipsum}                    
\usepackage{mwe}                       
\usepackage{amsmath} 
\usepackage{algorithm}
\usepackage{algpseudocode}
\usepackage{graphicx}
\usepackage{mdframed}
\usepackage{mathptmx}                  
\usepackage{array}
\usepackage{booktabs}
\usepackage{longtable}
\usepackage{appendix}

\begin{document}



\firstsection{Introduction}

\maketitle

Visual data storytelling is a powerful method for conveying a series of interconnected data facts and visualizations through a narrative flow. Among data-driven storytelling techniques, fact sheets are particularly effective. They integrate various data visualizations extracted from a dataset and weave them into a coherent narrative that effectively conveys key findings~\cite{shi2020calliope}. Fact sheets present information in a concise, single-page format, enhancing viewer retention and memorization~\cite{wang2019datashot}. With the digital transformation leading to an overwhelming influx of data, fact sheets have become increasingly vital for summarizing and presenting information.

Creating a fact sheet involves intricate tasks, from identifying compelling data points to designing visualizations and crafting a persuasive narrative~\cite{ryan2018visual}. Despite advancements in commercial tools like Excel and Power BI, which suggest appropriate visual encodings for datasets~\cite{becker2019microsoft}, substantial manual effort is still required. Tasks like identifying crucial data points and designing fact sheets demand considerable expertise in data analysis and graphic design. To address these challenges, research has introduced tools that streamline various aspects of fact sheet creation~\cite{wu2021ai4vis}. These tools automate the generation of visualizations from tabular datasets, simplifying the creation of compelling visuals~\cite{dibia2018data2vis}. Additionally, they facilitate fact sheet layout design, ensuring clear and effective information presentation~\cite{cui2019datasite, wang2019datashot}. By assisting in narrative composition, these tools reduce the learning curve for users without extensive data analysis expertise~\cite{shi2020calliope}. DataShot~\cite{wang2019datashot}, introduced in 2020, was the first tool to automatically generate fact sheets from tabular data, applying heuristics to rank and generate relevant facts. Calliope~\cite{shi2020calliope} further enhanced this by improving the relationships between facts for better content cohesion and logical flow.

However, existing approaches to automatic fact sheet generation struggle to fully capture the semantics of dataset and present them effectively. At the fact component level, these methods often fail to grasp both the meaning of individual data columns and the relationships between them, leading to missed insights and the generation of irrelevant conclusions. For instance, when analyzing the Tourism dataset (as shown in \cref{fig:teaser}), which includes columns for country names, GDP, and tourism revenue, these methods might inappropriately aggregate GDP values across countries, rather than correctly attributing the contribution of tourism to each specific country's economy. At the fact sheet level, these approaches struggle to establish relevance between facts and form a coherent narrative, resulting in a lack of interconnection and a disjointed flow of information. Additionally, the reliance on slot-filling, template-based methods for text generation leads to content that can be difficult for general readers to comprehend especially without providing related context of the dataset, such as unnatural statements like \emph{``receipts\_in\_billions of France is 35.96''}. These issues negatively impact the effectiveness of fact sheets, highlighting the need for more advanced techniques that make fact sheets accessible to a broader audience.

The rapid progress in generative AI has driven the development of Large Language Models (LLMs), such as GPT~\cite{openai_2023} and Llama~\cite{touvron2023llama}. These models, with in-context learning~\cite{min2022rethinking}, have demonstrated remarkable proficiency in performing logical reasoning tasks~\cite{chang2023survey}. Their inherent efficiency in handling diverse tasks without extensive retraining represents a promising approach to providing AI solutions across various domains. In data visualization, existing research has explored the utility of LLMs at the level of individual visualizations~\cite{wu2021ai4vis,chat2vis}. While these studies have shown improvements in segregated data analysis tasks, generating a fact sheet is a complex task involving multiple sub-steps, each requiring distinct expertise—from planning facts, extracting data, to drawing visualizations and composing the fact sheet layout. Recently, to enhance the capability of LLMs for complex tasks, researchers have explored AI chain design~\cite{handler2023taxonomy}, where it involves multiple AI-based worker with specialized expertise to excel in its modular task. This design has improved output quality across domains, including online discussions~\cite{dong2024multi} and social simulations~\cite{pang2024self}. Building on previous research, our primary objective is to investigate how AI-chain systems can improve the quality of generated fact sheets. Additionally, our research examines how these AI advancements can be integrated into data tasks, aiding not only data scientists but also novice users.

We introduce \tool, an automated fact sheet generation and customization tool designed to streamline the creation of fact sheets from raw datasets. \tool employs a novel approach featuring the collaboration of five specialized AI workers, each with distinct roles in processing and transforming the data. Upon receiving an input dataset and user requirements, \tool generates a dataset representation and passes it to the AI workers, where each worker executes its designated task, sequentially passing the output to the next worker in the pipeline. The result is a comprehensive fact sheet comprising interconnected facts. To enhance human-AI interaction, \tool includes a customization platform that allows users to review and modify the generated fact sheet using natural language commands, ensuring the final output aligns with specific user preferences.

To demonstrate the usability of the fact sheets generated by \tool, we present two fact sheets based on real-world datasets. Furthermore, we assessed the quality of our automatically generated fact sheets by conducting a user study with 18 participants. The results showed a preference for \tool's fact sheets over those from two state-of-the-art baselines~\cite{shi2020calliope, openai_2023}, citing more comprehensive content and superior visual appeal. Additionally, we validated the customisation interface by asking users to experiment \tool to customise a fact sheet and rate the quality of their customised fact sheets alongside their user experience. Participants confirmed that the customised fact sheets aligned with the dataset and their specific requests and praised the visual design. They also reported a positive user experience through the System Usability Scale questionnaire.

To summarise, the contributions of this paper include:
\begin{itemize}
\item We proposed a taxonomy for configuring AI chain workers and introduced a tool, \tool, that leverages this taxonomy for fact sheet generation. This demonstrates the applicability of AI chains in managing complex, multi-step workflows within the domain of data storytelling.
We hosted \tool as a public web-based automatic fact sheet generation and customisation platform for data community\footnote{\url{https://statement-driven-vis-15ebbabf813c.herokuapp.com/}}. 
\item We investigated the dataset representation aims for scalability and secure dataset communication to LLMs. Our data anomysation techniques propose the first step towards responsible AI usages for data works.
\item We used real-world datasets to illustrate the capability of our proposed system and validated \tool through a user study with 18 participants, which confirms its effectiveness and applicability across different expertise levels. The feedback provides future directions for human-AI collaboration on data tasks.
\end{itemize}

\section{Related Works}

\subsection{Data Analysis and Visualisation Tools}

Data analysis and visualization tools have seen significant advancements in recent years. Earlier tools like Excel and R required substantial manual effort and expertise, but there has been a notable shift towards automation. This transformation is driven by research efforts that have progressively automated various aspects of the data analysis process~\cite{wu2021ai4vis}. Initial research focused on providing visualization recommendations from datasets~\cite{cui2019datasite,kim2020gemini,wongsuphasawat2015voyager}, aiding users in the visualization process. Other work improved existing visualizations by focusing on aspects such as visual style searching~\cite{hoque2019searching}, assessing visualizations~\cite{fu2019visualization}, and annotation placement~\cite{bryan2016temporal}. Industrial tools like Tableau and Power BI have begun integrating AI to streamline data management for business applications~\cite{itsnotaboutthecell,nichols_wang_2024}. Although these systems have proven useful for data workers, novice users still face challenges in creating visualizations due to a lack of expertise.

Recent research has shifted towards automatically generating charts from datasets. Data2Vis~\cite{dibia2018data2vis} and VizSmith~\cite{vizsmith} demonstrated the feasibility of generating structured code in Vega-Lite and Python to produce appropriate graphs. Despite these advancements, a common limitation has been the inadequate integration of the human perspective in the outputs. The generated visualizations may not fully align with user intentions, limiting the usability of these systems. NL4DV~\cite{narechania2020nl4dv} addressed this by enabling chart generation based on natural language requests, aligning outputs more closely with user preferences. This capability was further enhanced by NL2Viz~\cite{wu2022nl2viz}, which incorporated program context to better capture user intentions. However, heuristic-based semantic matching in these tools still struggled with linguistic challenges, such as synonyms. For instance, terms like \emph{``yearly''} and \emph{``annual''} were not recognized as equivalent. Recently, LLMs have been applied to data visualization tasks, solving the problem of natural language semantic understanding and accommodating greater flexibility in user input~\cite{chat2vis}. While this research has shown positive results with various prompt techniques, we aim to explore how specialized AI worker definitions can further improve the quality of generated charts.

\subsection{Automatic Story-Telling and Fact Sheet Generations}

In scenarios where isolated visualizations fail to convey complex datasets effectively, narrative visualizations play a crucial role in presenting data with an intuitive and coherent flow, thereby enhancing accessibility for general audiences~\cite{ren2023re}. Narrative visualizations encompass a range of representations, including data videos~\cite{wang2021animated}, data comics, storylines~\cite{liu2013storyflow}, and infographics~\cite{cui2019text}. Early technologies provided platforms for users to manually design story flows. For example, Infonice~\cite{wang2018infonice} assisted users in creating infographics, bridging the gap between data exploration and presentation. Similarly, DataSelfie~\cite{kim2019dataselfie} enabled users to create personalized data visuals. However, these tools required a deep understanding of data and experience in composing data storytelling artifacts, limiting their accessibility to general users.

Recent technological advancements have led to the development of automated storytelling approaches~\cite{chen2023does}, offering a more accessible method for composing storytelling artifacts. These tools capture information from datasets and craft narratives around them. AutoClip~\cite{shi2021autoclips}, for example, automatically generates data videos to showcase fact findings. NewsViews~\cite{gao2014newsviews} provides an automated pipeline that extracts topics and creates corresponding geographic visualizations using contextual information from articles. Focusing on a more concise and data-driven format, recent advancements have also targeted the automated generation of fact sheets. DataShot~\cite{wang2019datashot} was the first tool to automatically generate fact sheets from tabular datasets, focusing on deriving quality facts from the data. However, this approach overlooked the overall story flow due to a lack of semantic relationship identification between different visualizations. Calliope~\cite{shi2020calliope}, an improvement upon DataShot, applied a logic-oriented Monte Carlo tree search algorithm to construct the story from facts. Despite this enhancement, Calliope still faces challenges in creating a seamless narrative across different visualizations due to limited understanding of data semantics. Additionally, previous fact sheet generation approaches did not consider user input and struggled with generating engaging textual components, such as chart titles and descriptions, due to their reliance on template-based methods~\cite{shi2020calliope, wang2019datashot}. Building on this evolving research trajectory, \tool aims to enhance the integration and natural flow of charts with more precise and natural captions.

\begin{figure*}[h]
  \centering 
  \includegraphics[width=\textwidth]{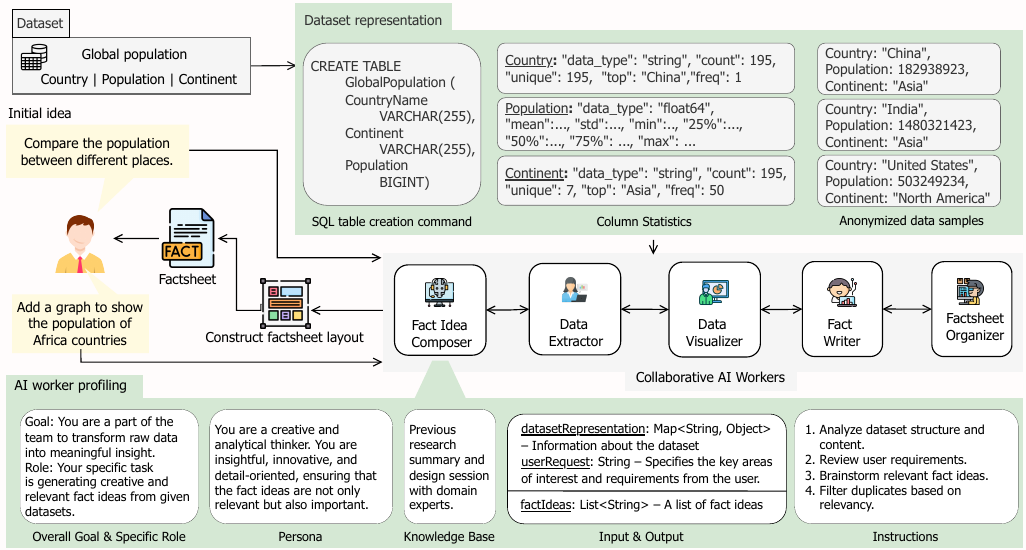}
  \caption{%
Workflow of \tool for generating a fact sheet using a global population dataset with columns for Country, Population, and Continent. \tool first prepares the dataset representation, which is then processed by the AI Chain. Each worker in the chain is profiled for specific tasks to generate the fact sheet.
}
  \label{fig:overview}
  \vspace{-4mm}
\end{figure*}

\subsection{Large Language Models for Data Tools}
Recent advancements in generative AI have driven the development of Large Language Models (LLMs), designed to understand natural language and generate human-like responses. Through prompt engineering techniques~\cite{min2022rethinking}, LLMs have improved the accuracy of heuristic-based solutions while reducing the need for extensive data and complex model training. Research on the applicability of LLMs spans various fields, from healthcare~\cite{cascella2023evaluating} and finance~\cite{wu2023bloomberggpt} to IT sectors like testing~\cite{deng2023pentestgpt}, automation~\cite{schwartz2023enhancing}, and smart devices~\cite{king2023get}. However, LLMs struggle with complex multi-step tasks, necessitating further research to enhance their usability. To address this, Wu et al. introduced AI chains~\cite{wu2022ai}, which break down complex tasks into manageable subtasks, each handled by a specific prompt. PromptSapper builds on this concept by defining the roles and collaboration of workers within the chain~\cite{cheng2024prompt}.

In the domain of data tools, pioneering research has explored the capabilities of LLMs in generating visualizations and aiding other visualization-related tasks. LLM4Vis, for instance, conducted experiments using a ChatGPT-based approach for visualization recommendations~\cite{wang2023llm4vis}, demonstrating the model's applicability for data tasks. Additionally, Chat2VIS~\cite{chat2vis} conducted a comparative study on the abilities of ChatGPT, Codex, and GPT-3 in rendering single visualizations from language queries. Other studies have also explored enhancing user interaction with visualizations by integrating LLMs to develop chatbot assistants for datasets~\cite{kavaz2023chatbot}. Recently, ChartGPT~\cite{tian2024chartgpt} performed fine-tuning of LLMs on a specialized dataset of visualizations to generate accurate and expressive charts from abstract natural language inputs. Additionally, DataTales~\cite{sultanum2023datatales} utilized LLMs' natural language capabilities to generate textual narratives accompanying charts in data-driven articles. This research has paved the way for advanced LLM applications in creating robust, user-centric data tools. Building on this, \tool adopts a holistic approach, leveraging LLMs to generate complete fact sheets from tabular datasets. Our goal is to apply AI chain techniques to tackle complex data tasks and validate their feasibility.

\section{System Design}

We introduce \tool, an automated fact sheet generation system, as shown in \cref{fig:overview}. \tool applies the concept of an AI chain to generate fact sheets based on data and user requests. Users can further customize the fact sheet, such as by adding new facts or rearranging content. In this section, we first highlight the AI worker profiling and workflow, followed by a detailed description of the fact sheet generation process. Lastly, we present the \tool’s interface for generating and customizing fact sheets.

\subsection{AI Worker Profiling \& Workflow}
\label{sec:agent}

While simple prompting techniques can effectively handle straightforward, isolated tasks, such as visualization code generation, they face challenges when applied to more complex, integrated tasks. First, LLMs often lack domain-specific knowledge, which can result in outputs that do not meet the nuanced requirements of specialized tasks. Second, providing an abstract task without thorough step-by-step planning can disrupt the overall process, as LLMs may fail to propose the necessary sequence of actions to achieve the desired outcome. Finally, there can be incompatibilities between subtasks, where outputs from one stage may not align with the inputs required by the next, leading to pipeline failures and hindering task completion. To address these issues, we adopted the AI chain approach for generating fact sheets from tabular datasets. This approach leverages specialized AI workers, known as agents, each designed to handle specific subtasks more efficiently. These agents possess unique knowledge and capabilities tailored to their specific tasks, allowing them to focus on their assigned roles with greater precision, thereby improving the overall quality of the output. We define each worker with the following attributes:

\textit{Overall Goal \& Specific Role.} To simulate the positive impact on work outcomes and productivity, we provide overall goals before detailing individual tasks, mirroring real-world scenarios~\cite{barrick2013theory}. The workers share common objectives while working on specific parts of the pipeline, stimulating goal-directed behaviors~\cite{sioutis2006agent}. The specific role of each worker in the AI chain is then defined to achieve the overall objective.

\textit{Persona.} Persona refers to how each individual behaves and thinks~\cite{10.1145/3613904.3642406}, which has been shown to be crucial in optimizing the effectiveness of AI workers~\cite{10.1145/3640794.3665887}. While LLMs do not have a default persona, we manually define the persona of each worker, including the characteristic traits, motivation, and expertise necessary for successfully completing assigned tasks. For example, the Fact Composer is specified to be creative and analytical, while the Data Extractor is meticulous and knowledgeable. Further details of the AI workers we defined can be found in Section~\ref{sec:factideas}.

\textit{Input \& Output.} We provide a list of input and output variables with their descriptions, formats, and data types that the agent receives or returns. This improves the reliability of the overall pipeline. Additionally, we include few-shot examples following this input and output format, which further enhances the reliability and alignment of the output~\cite{song2023llm}.

\textit{Knowledge Base.} This refers to specialized domain knowledge that supplements the general capabilities of LLMs. For each task assigned to a worker, we compile and refine relevant datasets and prior research to build a tailored Knowledge Base. This allows the agent to perform optimally in its specialized task, avoiding the hallucination effects of LLMs when dealing with complex user requests~\cite{martino2023knowledge}.

\textit{Instructions.} Research has shown the importance of sub-step guidance over abstract instruction for task performance accuracy~\cite{madaan2022language}. Therefore, we include step-by-step procedures that each worker follows to achieve its objectives.

Our AI workers utilize a decentralized communication model, where interaction occurs directly between related agents~\cite{cheng2024prompt}. Upon completing a task, an agent forwards its output to the next designated agent in the workflow. To maintain consistency, all input and output data are standardized in JSON format, adhering to REST API protocols. For transmitting object files, such as images or data files, these are stored as block objects, with the directory path provided for file retrieval.

\subsection{Fact Sheet Generation}

Upon receiving the dataset and, optionally, a user request, the fact sheet generation process begins. \tool first analyzes the dataset to create a dataset representation and then utilizes the AI workers described in \cref{sec:factideas} to generate the fact sheet structure and visual components. Following this, we arrange the fact sheet contents on the canvas and export the initial fact sheet.

\subsubsection{Dataset Representation}
\label{sec:anomysation}

Given that LLMs process input as textual prompts, our goal is to identify the optimal method for capturing and conveying the essential details of the input dataset. Traditional approaches often involve feeding the entire dataset to the LLM, but this method faces significant limitations. Firstly, LLMs have token constraints, making it impractical to handle large datasets. Additionally, directly exposing the entire dataset can compromise privacy and intellectual property. Therefore, our approach focuses on incorporating sufficient dataset information into the prompts to guide LLMs in making accurate decisions without directly providing the dataset. We achieve this by presenting the dataset using SQL table creation commands, column statistics, and example rows. An example of dataset representation for the Global Population dataset with 3 columns (Country, Population, Continent) is shown in \cref{fig:overview}.

\textit{SQL Table Creation Command.} To capture the table name and column names along with their corresponding data types, we use SQL table creation command syntax. This approach also aids our AI workers in better formulating and generating SQL query syntax in subsequent processes. 

\textit{Column Statistics.} For each column, we generate statistics that summarize the values it contains. For numerical columns, this includes the maximum, minimum, median, average, and the 25th and 75th percentiles. For string columns, we count the number of unique values and identify the top five most frequent occurrences. These statistics provide LLMs with a comprehensive overview of the data range.

\textit{Anonymized Data Examples.} We include anonymized sample rows from the dataset to help LLMs understand the expected data values in each column. The number of rows presented depends on the token limits, with more columns requiring more tokens. Given that the input data may contain private information or intellectual property, anonymization is crucial to protect the confidentiality of the data~\cite{lin2024promptcrypt}. We employ a format-preserving encryption technique that maintains semantic integrity, preserving the semantic meaning and ordinal relationships of the data during anonymization. This allows LLMs to comprehend the underlying data structure without directly accessing the original dataset. The anonymization process involves classifying each column by its data type—nominal, ordinal, discrete, or continuous—and applying a tailored anonymization method.

\begin{itemize}

\item \textit{Nominal values} are qualitative and lack a specific order or comparability. To anonymize these values while preserving meaning, we substitute them with semantically equivalent terms. For instance, if the column refers to country names, ``France'' might be replaced by ``Italy.'' We use the column's entity type to guide this substitution, generating random values for each entity type and mapping them to unique values in the data.
\item \textit{Ordinal values} have a specific order, and maintaining this order during substitution is crucial to avoid inconsistencies in later analysis. For example, in a letter grading system where grades are A, B, C, D, and F, substituting these grades with random values could lead to confusion, such as when querying how many students received a grade better than C. To prevent this, we first extract all valid values from the dataset and ensure that the anonymized values are selected from this pool.
\item \textit{Discrete numerical values} are anonymized using a mathematical transformation that maintains the ordinal relationships between values. For example, if the original values are integers representing counts, the transformed values will still reflect the same relative magnitudes, ensuring consistency in any subsequent analysis.
\item \textit{Continuous values} such as those found in time series data, are treated with a different approach. We anonymize these by generating new random values within the original range, ensuring that the data remains realistic and usable while masking the exact figures.
\end{itemize}

\subsubsection{Fact Sheet Content Generation}
\label{sec:factideas}

After anonymizing the dataset, we feed the data into our AI chain to generate the fact sheet content. Users can provide additional requests in natural language to specify their requirements for the dataset. This feature is particularly useful when users are interested in covering only a specific portion of the dataset rather than the entire dataset. For example, a user may request data within a limited time range or focus on specific columns. This section discusses how each agent handles its tasks in the pipeline.

\textit{Fact Idea Composer.} 
The primary task of this agent is to generate fact ideas from the dataset. To enhance the agent's capability, we incorporated expert knowledge from a prior study that categorizes facts into eleven distinct types such as Distribution, Categorization, Trend and Rank~\cite{wang2019datashot}. Additionally, we conducted a design session with three senior data scientists, providing each with ten distinct datasets and asking them to suggest interesting facts for a fact sheet. We refined these suggestions by removing duplicates and resolving ambiguities, integrating the refined list of fact ideas into the Fact Idea Composer's knowledge base\footnote{\url{https://anonymous.4open.science/r/factflow-public-2BF8/knowledgeBase}}. This ensures that the generated fact ideas align closely with expert expectations and are grounded in the actual data context. The agent is instructed to generate all potential facts and rank them based on relevance and significance, filtering out less important or redundant facts. The Fact Composer Agent employs the self-consistency prompting technique~\cite{cheng2024relic}, using multi-turn communication to ensure output accuracy. The agent's output consists of a list of fact ideas, each categorized by fact type and content.

\textit{Data Extractor.}
Since each fact represents only a portion of the original dataset, \tool involves the Data Extractor to retrieve the data relevant to the fact. This agent receives a fact idea, additional user requests, and the dataset representation to generate an appropriate SQL command to query the original dataset. To improve the accuracy of converting natural language to SQL (NL2SQL), we added into the agent's knowledge base the Spider dataset~\cite{yu-etal-2018-spider}, which is a well-known dataset containing the natural language questions and its corresponding SQL command. Given the challenges LLMs face in code generation, particularly in terms of robustness and accuracy~\cite{yu2023llm}, we apply the Code Generation with Advising and Validation approach~\cite{chen2023seed}. The agent first lists recommendations for code generation, considering data types and required facts. It then generates an executable SQL command to query the data locally. In the validation step, the agent presents the data representation obtained from the initial query, along with the SQL command, to the LLMs to refine any syntactical errors. Upon completion, the agent outputs the filtered dataset relevant to the fact idea.

\textit{Data Visualizer.}  
After filtering the data, each fact idea is presented through a visualization. The Data Visualizer uses the dimensions of the filtered dataset provided by the previous agent to recommend the most appropriate visualisation type and visual encoding channels  To ensure the fact sheet's accessibility to a wide range of users, \tool supports widely recognized visualization types such as line charts, bar charts, scatter plots, pie charts, and area plots. We limit the number of encoded dimensions to three, resulting in a 2D chart with color as the maximum additional encoding. Unlike previous approaches that relied on direct code generation for creating graphs, we implemented a parameter-driven approach. In this method, a function generates different types of visualizations based on input parameters provided by the agent. These parameters may include chart type, data dimensions, axis labels, color schemes, and more. To enhance the Data Visualizer's capability in generating accurate and relevant visualizations, we documented the function’s input and output parameters. This documentation serves as a knowledge base, guiding the agent in appropriately configuring the parameters to produce the desired visual outputs.

\textit{Fact Writer.}
The Fact Writer leverages the processed dataset and visualizations from previous agents to generate factual statements that are well-aligned with the context provided by the charts. Additionally, the agent aims to clarify the underlying rationale of the presented data. While some inquiries related to the data can be directly addressed using the processed data from previous steps, others require a broader knowledge base for comprehensive explanation. Therefore, we instruct the agent to generate questions regarding the causal relationships in the presented data. The Fact Writer uses LLMs to create additional questions that explore the causation effects of external contexts, such as related historical events or geographical information. These questions help explain the key data points highlighted in the chart or explain what the fact implies. For example, if the dataset shows that France is the most visited country for tourism, the agent might ask, \textit{``Why is France the most visited country for tourism?''} to gain deeper insights and present them. This enriched content significantly enhances the viewer's understanding and interpretation of the charts, especially when dealing with complex data that might be unfamiliar to the audience.

\textit{Factsheet Organizer.}
The primary goal of this agent is to structure the fact sheet in the most logical order to enhance the learnability of the presented information. Our fact sheet contains different sections, each presenting a key topic with a list of relevant charts to support the idea. Using the list of facts and generated material from other workers, the Fact Sheet Organizer finds relevancy between facts to form the story flow with several key topics. After that, the agent sorts the list of facts into these topics based on the relevancy and outputs the structure of the fact sheet in JSON format for fact sheet generation.

\begin{algorithm}
\caption{Split Sections Into Columns}
\label{alg:split_sections}
\begin{algorithmic}[1]
\Require A list of $sections$
\Ensure A boolean list for column layout
\Function{split\_columns}{$sections$}
    \State $best\_diff \gets \infty$, $best\_layout \gets \text{None}$
    \For{$perm$ in $\text{permutations}(sections)$}
        \For{$comb$ in $\text{powerset}(\text{range}(1, \text{len}(perm)))$}
            \State $left\_score \gets \text{calculate\_score}(perm[0])$
            \State $right\_score \gets 0$
            \For{$i \gets 1$ \textbf{to} $\text{len}(perm) - 1$}
                \If{$i$ in $comb$}
                    \State $left\_score \gets left\_score + \text{calculate\_score}(perm[i])$
                \Else
                    \State $right\_score \gets right\_score + \text{calculate\_score}(perm[i])$
                \EndIf
            \EndFor
            \State $diff \gets \text{abs}(left\_score - right\_score)$
            \If{$diff < best\_diff$}
                \State $best\_diff \gets diff$, $best\_layout \gets (perm, comb)$
            \EndIf
        \EndFor
    \EndFor
    \State Convert $best\_layout$ to boolean list $result$
    \State \Return $result$
\EndFunction
\end{algorithmic}
\end{algorithm}

\subsubsection{Fact Sheet Layout}
\label{sec:layout}

\begin{figure*}
  \centering 
  \includegraphics[width=0.8\linewidth]{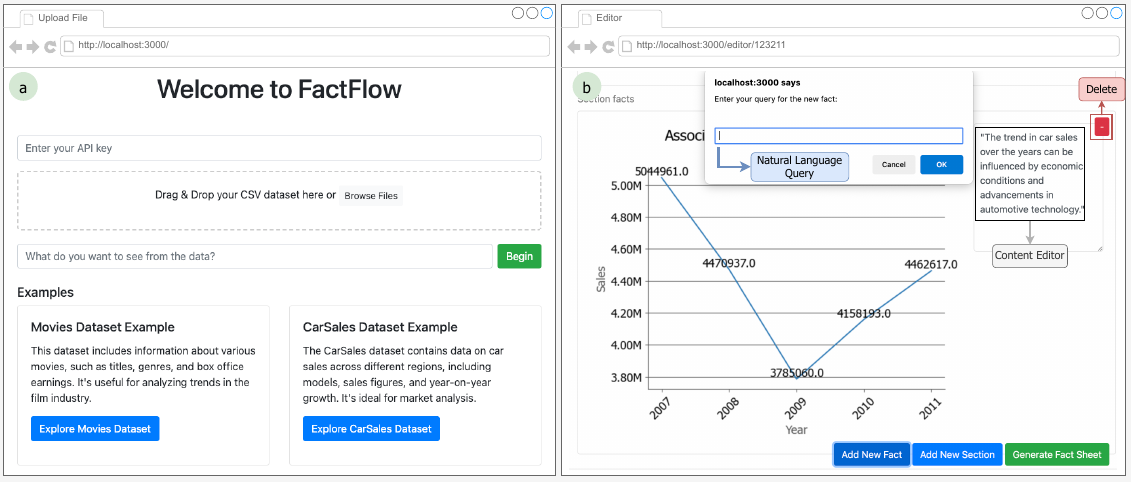}
  \caption{%
  	Our \tool interface: a) User can upload dataset and provide any additional natural language request to generate the fact sheet b) Based on the initial fact sheet, users can manually add, remove and edit the content, as well as generating a new fact using natural language request.
  }
  \label{fig:customisation}
    \vspace{-4mm}
\end{figure*}

After establishing the structure of the fact sheet, we proceed to design its layout. We advocate for a dual-column layout to enhance readability and minimize eye movement~\cite{shrestha2008eye}. The layout consists of multiple sections within each column, arranged vertically. Our objectives in developing the layout are twofold: 
First, to preserve the original logical order of the sections as outlined in the fact sheet structure, thereby enhancing its storytelling attributes. Second, to achieve a balanced distribution of content between the two columns. This balance not only contributes to the aesthetic appeal of the fact sheet but also ensures content readability. A key challenge is dividing sections into two columns, given the varying sizes of each section due to differing numbers of facts. To address this, we assign a fixed height to each fact and calculate the cumulative height of each section. We then apply an optimization algorithm to distribute sections across the columns, as described in \cref{alg:split_sections}. This algorithm processes an array of section heights in logical order and outputs a boolean array, assigning sections to the left column as ``True'' and to the right column as ``False''. The Introduction section is always placed at the top left, so the first array value is set to ``True'' by default. \tool then uses this layout structure to iteratively assign sections to the columns, generating the complete fact sheet.

\subsection{Fact Sheet Customization} 
\label{sec:customisation}

In developing our automated pipeline for fact sheet generation, we emphasize the synergy between human input and AI automation, aiming to enhance human-AI collaboration with AI as an assistant in various tasks. To support this, we provide an interface that allows users to create and customize AI-generated content within the fact sheet. Users start by uploading their dataset on the Upload page (\cref{fig:customisation}a) and can specify their interests. We also offer two sample datasets for experimentation.

Upon clicking Begin, \tool automates the fact sheet generation and directs users to the Editor page (\cref{fig:customisation}b), where they can dynamically tailor the content to their needs. The interface allows users to add, remove, and rearrange sections according to their preferences. For each visualisation, users can delete, rearrange or move them across sections. \tool supports adding new charts based on natural language request, processing this as new fact ideas through the AI chain to generate corresponding charts and textual content. This new fact will be presented in appropriate section automatically. After customization, users can export the fact sheet as a PDF. The generated content will be stored with the unique ID, where user can further continue to modify the same fact sheet in the future.

\subsection{Implementation}

\begin{figure*}[!h]
  \centering
   \label{fig:showcase}
  \begin{subfigure}[t]{0.50\textwidth}
  	\centering
  	\includegraphics[width=\textwidth]{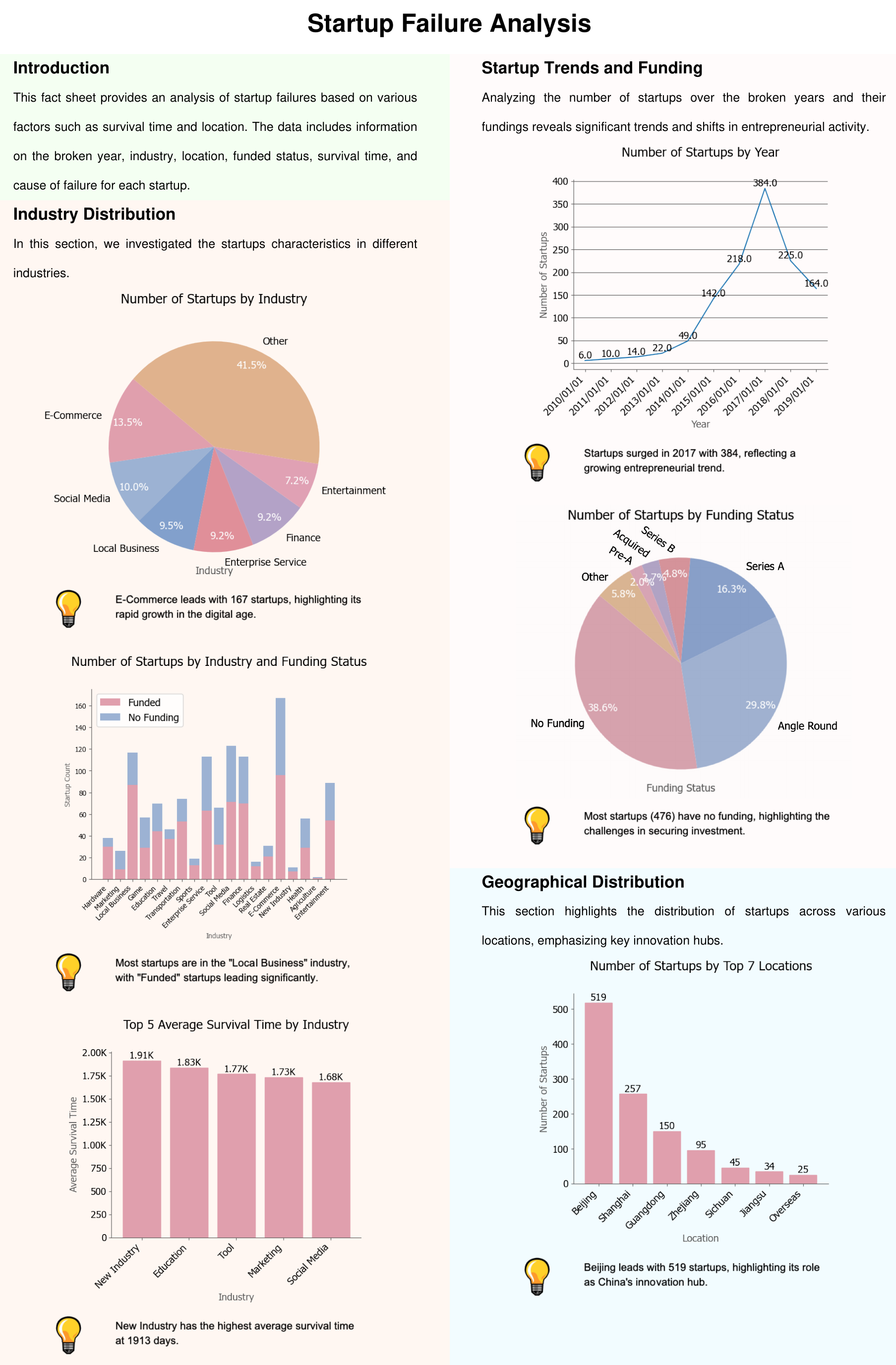}
  	\caption{Fact sheet about startup failures after the tide of ``new economics'' in China.}
  	\label{fig:showcase1}
  \end{subfigure}%
  \begin{subfigure}[t]{0.49\textwidth}
  	\centering
  	\includegraphics[width=\textwidth]{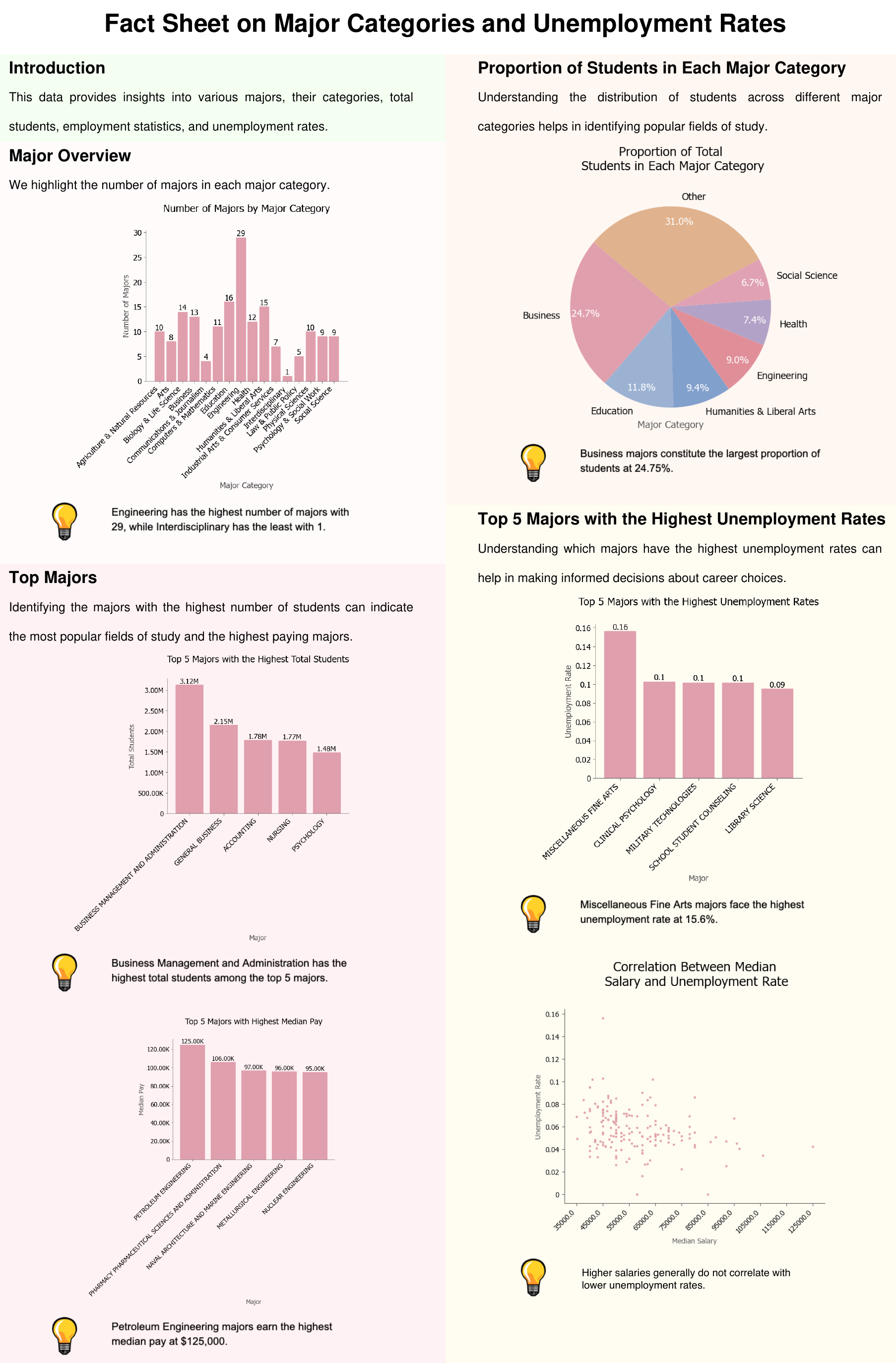}
  	\caption{Fact sheet about university major and employment opportunities.}
  	\label{fig:showcase2}
  \end{subfigure}%
  \\%

  \subfigsCaption{Two fact sheets that automatically generated by \tool based on real-world datasets. }
    \vspace{-4mm}
\end{figure*}

\tool is implemented in Python 3.8, utilizing the Matplotlib library for generating visualizations and the FPDF library for creating and storing fact sheets. To facilitate seamless communication with the GPT-4 model from OpenAI, we establish a REST API interface. For public accessibility, \tool is deployed as a web application using Flask.

\section{Evaluation}

We evaluate the usefulness of \tool through (1) two data stories generated by \tool to showcase the quality of the fact sheets, and (2) a comprehensive user study to validate both the quality of the generated fact sheets and the usability of the customization module.

\subsection{Use Cases}

To demonstrate the capabilities of our automatic generation tool, we present two illustrative fact sheets in \cref{fig:showcase}. These fact sheets distill complex datasets into clear, insightful narratives using visualizations and storytelling.

\Cref{fig:showcase1} presents a dataset about Startup Failures, featuring 1,234 rows and 6 columns that record the details of companies that closed during or after the ``new economics'' boom in China from 2010 to 2019. This dataset covers various aspects of startup profiles, including the broken year (e.g., 1/1/2019), industry (e.g., Finance), location (e.g., Shanghai), funding status (e.g., Series B), survival time in days (e.g., 1,730), and cause of failure (e.g., supervision). The initial request was \emph{``Analyze the reasons for startup failures across different locations and industries.''} \tool structured the analysis into sections: industry distribution, startup trends and funding, and geographical distribution. The Industry Distribution section reveals the dominance of the E-Commerce sector, while noting that the Education industry has the highest average survival time. The Startup Trends and Funding section highlights a peak in startups in 2017 and notes that most lack funding, underscoring investment challenges in various industries.  The Geographical Distribution section identifies Beijing as a major innovation hub, leading with 519 startups, reflecting its significance in China's startup ecosystem. 

\Cref{fig:showcase2} features a dataset on university majors and employment opportunities, encompassing 173 rows and 12 columns. This dataset includes statistics on various university majors, such as the major's name (e.g., Accounting), its broader category (e.g., Business), total student enrollment (e.g., 1,778,219), employability statistics (e.g., Employed: 1,335,825; Unemployed: 75,379; Unemployment Rate: 0.07), and the median pay (e.g., \$65,000) associated with each major. The initial request was \emph{``Facts about university majors and employment status.''} \tool structured the analysis into four sections: Education Majors Overview, Top Majors, Proportion of Students in Each Major Category, and Top 5 Majors with the Highest Unemployment Rates. The Education Majors Overview notes that the Engineering category has the highest number of distinct fields, with 29 different majors. It then identifies the most popular majors, with Business Management and Administration leading in student enrollment among the top five. In terms of earning potential, Petroleum Engineering majors stand out with the highest median salary of \$125,000. The analysis continues with the distribution of students across various major categories, highlighting that Business majors constitute the largest proportion at 24.75\%. Lastly, the section on unemployment rates reveals that Miscellaneous Fine Arts has the highest unemployment rate at 15.6\% and displays the correlation between the median salary and unemployment rate.

\subsection{User Evaluation}

We conducted a user study to assess the quality of the fact sheets generated by \tool and the usability of our fact sheet customization interface. First, to evaluate the quality of the fact sheets created by \tool, we involved data workers from industry and non-data workers to determine the tool's versatility in catering to both data professionals and general audiences. We presented the automatically generated fact sheets from \tool and two state-of-the-art baselines to the participants and collected their feedback. Second, we assessed the quality of customized fact sheets and the user experience during the customization process using \tool. This involved a user study where participants actively customized their fact sheets with \tool, followed by feedback collection.

\subsubsection{Research Questions}
This study aimed to answer the following research questions:
\begin{itemize}
    \item \textbf{RQ1}: How does the content and visual presentation of fact sheets generated by \tool compare to those produced by existing approaches?
    \item \textbf{RQ2}: How does a user's experience level with data analytics influence their evaluation of fact sheets?
    \item \textbf{RQ3}: What is the quality of customized fact sheets created using \tool, and how do users perceive the experience of customizing fact sheets with this tool?
\end{itemize}

    \begin{figure*}[tb]
  \centering 
  \includegraphics[width=0.95\linewidth]{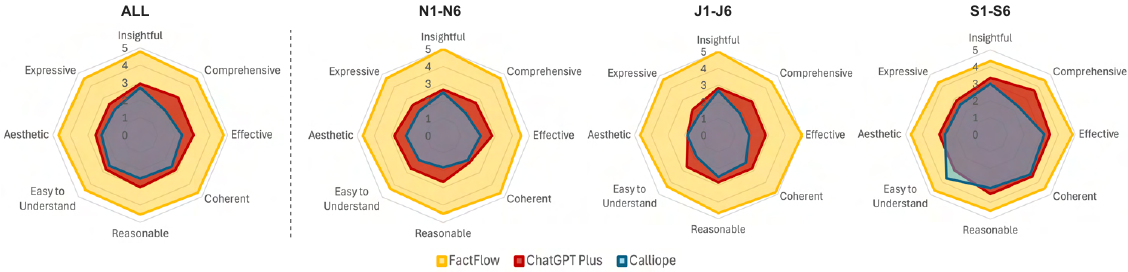}
    \caption{Radar charts showing the quality ratings of fact sheets produced by \tool compared to the Calliope and ChatGPT Plus baselines, evaluated by three participant groups: N1-N6 (novices), J1-J6 (junior-level data workers), S1-S6 (senior data workers), and ALL (all participants).}
  \label{fig:part1}
    \vspace{-4mm}
\end{figure*}

\subsubsection{Experimental Setup}

    \textbf{Participants.} We recruited eighteen participants (8 males, 10 females, aged 23-46 years, Mean=30.94, SD=1.43) to join our 2-phase study. All were proficient English speakers with at least a bachelor's degree. Participants were divided into three groups based on their experience with data visualizations: Group 1 (N1-N6) with no prior experience with data work, Group 2 (J1-J6) with 1 to 5 years of work experience, and Group 3 (S1-S6) with over 5 years of experience. This division aimed to analyze the impact of work experience on expectations of fact sheet quality and the user experience with \tool. Each participant received a AU\$50 Amazon gift card for their participation.

    \textbf{Datasets.} We used two different datasets to generate fact sheets for this experiment. First, we selected the CarSales dataset (as used in previous research~\cite{shi2020calliope}), comprising 4 columns (Brand, Type, Sale, and Year) and 275 rows, with each row representing a car sale record. Second, we used the Movies dataset, which includes 7 columns (Movie, Studio, Type, Worldwide \$m, Domestic \$m, Overseas \$m, and Year) and 198 rows, with each row representing a movie. This dataset was chosen for its diversity in data types (categorical, temporal, and numerical), enabling the extraction of a wide range of facts. All participants were unfamiliar with these datasets before the user study.

    \textbf{Baselines.} We selected two baselines that support automatic fact sheet generation and generated two fact sheets each for the CarSales and Movies datasets. The first baseline, \textbf{Calliope}~\cite{shi2020calliope}, is a state-of-the-art tool for producing high-quality narrative flows in fact sheets. We used Calliope to benchmark improvements in fact sheet generation. The second baseline, \textbf{ChatGPT Plus} with the latest GPT-4 model, is recognized for its capability to interpret data files and generate graphical visualizations~\cite{openai_2023}. We used ChatGPT Plus to demonstrate the advancements of our AI chain approach compared to simple prompting techniques. We uploaded the datasets to ChatGPT Plus and requested fact sheets be generated. Fact sheets from \tool were generated using our public platform. Each fact sheet was automatically generated with no further modification or changes. All generated fact sheets from \tool and the baselines are provided in our supplementary material\footnote{\url{https://anonymous.4open.science/r/factflow-public-2BF8/Experiment/}}.

    \textbf{Procedures.} We conducted a 60-minute Zoom session with each participant, initially introducing them to the datasets and the concept of fact sheets. After this introduction, we displayed six automatically generated fact sheets—two each from \tool, Calliope, and ChatGPT Plus, derived from the Movies and CarSales datasets. Participants were given five minutes to review each fact sheet. To maintain fairness, the presentation order of the fact sheets was counterbalanced, and participants were not informed which tool produced which fact sheet. We then collected both quantitative and qualitative feedback for each fact sheet, ensuring an unbiased comparison and comprehensive evaluation.

    After collecting feedback on the generated fact sheets, we introduced participants to our customization platform through a 3-minute tutorial video. This video used the CarSales dataset to showcase \tool's capabilities, such as adding or deleting facts and sections, rearranging components, and editing text as detailed in \cref{sec:customisation}. Following the tutorial, participants were given a \tool-generated preliminary fact sheet based on the Movies dataset, with 20 minutes allocated for customization according to their preferences and dataset insights. Subsequently, we collected both quantitative and qualitative feedback through a structured questionnaire.

    \textbf{Metrics.} Similar to Calliope's evaluation~\cite{shi2020calliope}, we assessed fact sheet quality based on Content (Insightfulness, Comprehensiveness, Effectiveness, Coherence, and Reasonableness) and Visuals (Ease of understanding, Aesthetic appeal, and Expressiveness). We removed the System aspects as we only evaluated the quality of generated fact sheets. For each metric, we used a 5-point Likert scale (1 being the lowest, 5 being the highest). Additionally, participants were asked to rate their most preferred fact sheet and justify their choice, as well as provide suggestions for improvement for each fact sheet.
    
    We assessed the quality of our \tool customization platform using three metrics: (i) \textbf{Accuracy}, which measures how well the generated facts reflect the dataset's data; (ii) \textbf{Alignment}, which assesses the extent to which the customized facts match the users' ideas and requests; and (iii) \textbf{Aesthetic}, which evaluates the visual appeal and design quality of the fact sheets. Participants rated each metric on a 5-point Likert scale (1 being the lowest, 5 being the highest). To measure the usability of \tool, we used the System Usability Scale (SUS)~\cite{bangor2008empirical}, which consists of 10 questions each rated on a 5-point scale (1 for strongly disagree, 5 for strongly agree). Finally, we invited participants to share their thoughts on the advantages and disadvantages of using \tool, their experiences with AI-assisted customization, and how they see such tools fitting into their daily data workflows.

\begin{figure*}[tb]
  \centering 
\includegraphics[width=0.8\textwidth]{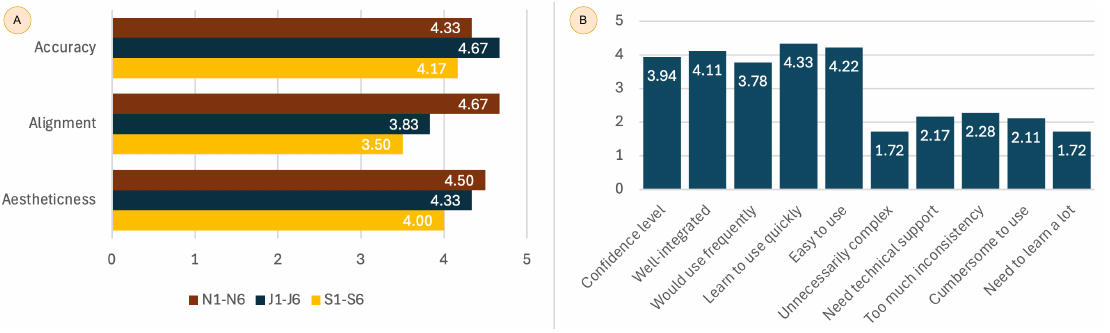}
  	\caption{A) Bar chart showing ratings from different user groups on Accuracy, Alignment, Aestheticness for the customized fact sheet.
   B) Bar chart displaying average SUS feedback from participants on the use of the \tool platform.}

  \label{fig:part2}
    \vspace{-7mm}
\end{figure*}
 
\subsubsection{Results}
We summarize feedback from 18 participants to answer the research questions, starting with quantitative assessments of \tool and baseline tools, followed by user preferences for fact sheets and their reasons. We also discuss suggestions to enhance \tool's fact sheet quality. After that, we detail user evaluations of the customized fact sheets and their experiences with \tool, alongside a discussion on the potential benefits of AI-assisted tools in data workflow optimization.

\underline{Fact Sheet Quality Ratings:} As depicted in \cref{fig:part1}, our quantitative analysis of participant feedback highlights \tool's impressive performance, with an overall average rating of 4.63 out of 5. This rating is significantly higher than the scores attained by Calliope and ChatGPT Plus, which received average ratings of 2.37 and 2.84, respectively. Remarkably, \tool surpassed these tools in every user group category, including users with no experience (N1-N6), junior data workers (J1-J6), and senior data workers (S1-S6). Our tool demonstrated particularly strong performance in the criteria of being \emph{Insightful} (4.78) and \emph{Effective} (4.78). The comparison between the ratings from different participant groups for \tool revealed that junior data workers gave it the highest rating of 4.71. This was closely followed by users with no prior experience, who rated it at 4.63, and senior data workers, who provided a slightly lower rating of 4.54.

\underline{Most Preferred Fact Sheet and Rationale:} The collected feedback indicates a strong preference for \tool among all participants, who commended its comprehensive and logical presentation of data. Participants N1, N5, N6, J1, and J2 were impressed by the coherent story that \tool narrated, with N1 describing it as \emph{``the most comprehensive one''} and N6 noting its storytelling capability that allowed them to have a deep understanding of the dataset with minimal effort. In terms of design and layout, participants praised its consistency and structure (N3, J4, J5). N3 admired \emph{``the consistency of fonts, colors, and visualization size,''} while J4 and J5 acknowledged its structured layout, with J5 noting that the consistency of layout helped to improve the overall readability of the fact sheet. Visual appeal and clarity were also highlighted as significant strengths (S1, S5, S6). S5 mentioned that it \emph{``uses more appropriate chart types and is more visually appealing,''} reinforcing S1's sentiment, who found that the \emph{``layout and design of the visualization are visually appealing.''} Furthermore, several participants, such as S3, suggested \tool's potential role in real-world applications, especially in explaining business-related data. The clarity with which \tool delivers information was emphasized by S4 as \emph{``understandable and very clear,''} highlighting the fact sheet's effectiveness in communication. Collectively, the feedback illustrates that \tool excels in providing a logical, comprehensive, and visually appealing representation of data, structured in a way that is both engaging and accessible for various users, from business analysts to those less skilled in data analysis. 

For the baseline tools, participants mentioned that they appreciated the color palette and layout of different facts from Calliope. For improvement, they suggested introducing diverse visual elements to comprehensively cover necessary information, optimizing chart types and designs for clarity, and ensuring a professional and uncluttered layout. Additionally, the language content can be utilized to provide causal explanations for data trends. For ChatGPT Plus's baseline, while the majority of participants praised the content for being easy to understand, the focus for further improvements is on visualizing key information succinctly, simplifying data representation, and generating deeper insights, possibly through more advanced analysis of the provided data.

\begin{mdframed}[linecolor=black, linewidth=1pt]
    \textbf{Answer to RQ1}: \tool outperforms baseline methods in generating fact sheet content that is not only more insightful and comprehensive but also more coherent and logically sound. Additionally, the tool excels in creating designs that are easier to comprehend and more aesthetically appealing and expressive.
\end{mdframed}

 \underline{Improvements:} To enhance \tool, the feedback indicates a need for brevity and visual clarity, with participants suggesting more concise content for quicker comprehension (N1, N2, N4). Additionally, participants advocated for more diverse visual elements, such as additional chart types and images, to increase engagement (N3, N6, J4). Some critiques focused on improving visual appearance through larger text sizes and distinct color use to differentiate data points (J6, S6). Finally, S2 and S3 suggested more intricate data representation, including forecasting and statistical analysis, from their experience.

\underline{Fact Sheet Customization Quality.} As shown in \cref{fig:part2}a, \tool received an average rating of 4.11 out of 5, reflecting robust user satisfaction. \tool scored 4.39 for Accuracy, indicating the robustness of our approach to query and present data from the dataset. The Alignment and Aesthetic Appeal received scores of 3.83 and 4.11, respectively, suggesting that \tool effectively resonates with user needs and aesthetic preferences. When analyzing the differences in results from different participant groups, all groups agreed on the aesthetics and accuracy of the modified fact sheet. Interestingly, there were differences in ratings regarding alignment for these user groups. By analyzing the commands issued by users to request new facts, we noticed that more experienced participants tended to have more complex requests. Users with low data experience made simple queries such as \emph{``Show me the top 5 dramas with the highest revenue this century''} or \emph{``The proportion of movies by type from Fox Studio,''} which \tool handled effectively. However, a request from a senior data scientist, \emph{``Predict the future trends of different studios in choosing the most profitable movie genre,''} posed challenges to \tool in terms of predictive modeling capabilities for forecasting future trends, which can be considered as an area for future work.

\begin{mdframed}[linecolor=black, linewidth=1pt]
    \textbf{Answer to RQ2}: The experience level of a participant is proportional to their expectations of fact sheet quality, as well as the complexity of their requests.
\end{mdframed}

    \underline{SUS Form Result.} \cref{fig:part2}b shows the usability feedback of \tool, resulting in a score of 73.45 out of 100. Specific feedback highlighted \tool's intuitiveness, with scores of 4.22 for ease of use and 4.33 for the ability to learn quickly. The integration quality received a positive 4.11, while complexity was deemed low, with a score of 1.72. Collectively, these metrics underscore the high usability of \tool.

    \underline{Qualitative Analysis.} We inquired about participants' most favored aspects of \tool and their recommendations for user experience improvements. Participant S1 and J1 praised \tool's innovative approach, while N2 highlighted the novelty of creating charts via commands as \textit{``quite new.''} Additionally, N4 appreciated the tool's performance, noting its \textit{``fast and responsive''} nature. Features such as the drag-and-drop functionality and PDF export capability received positive feedback for their usefulness in data sharing and presentation, with N6 emphasizing their value. To enhance user support, suggestions were made for integrating a chatbot for immediate assistance (N1) and providing on-screen tutorials to facilitate a smoother onboarding experience (N3). An undo feature was proposed by S6 to offer more control over the data visualization process. Meanwhile, N4 expressed a desire for customizable formats to suit various presentation contexts.

    \begin{mdframed}[linecolor=black, linewidth=1pt]
        \textbf{Answer to RQ3}: The customization module of \tool received high ratings for usability and output quality, showcasing its effective customization capabilities.
    \end{mdframed}

\section{Discussion}




\textbf{Modern Data Analysis Tools and Privacy Considerations in AI- Enhanced Workflows.} 
The rapid evolution of AI has significantly reshaped the data analysis and presentation process. From traditional, fully manual tools that required extensive effort, research has increasingly focused on making the data analysis process more intuitive through automation, reducing user effort while maintaining output quality~\cite{grudin2009ai}. While AI automation offers great benefits by reducing repetitive tasks, fully automated systems that exclude human involvement are often not intuitive. Previous research~\cite{li2023ai} shows that data workers are hesitant to adopt AI tools due to its inherent weaknesses, such as limited creative capabilities and a lack of understanding of data context. The key challenge is finding a balance between human expertise and AI-driven automation, with some tools requiring substantial manual operations and others overly automating the process without adequate user oversight~\cite{wu2021ai4vis}. From our evaluation, the user feedback highlights that while AI effectively generates preliminary insights and simplifies data processing, human refinement is still essential to tailor results to specific needs. As AI tools advance, their role as interactive assistants—providing suggestions and answering queries—represents a promising direction. However, the integration of AI, particularly LLMs, into data workflows has raised privacy concerns. Skepticism about sharing private dataset with AI systems underscores the importance of securing private information to maintain trust in AI technologies~\cite{Gal,9152761,brown2022does,Gurman_2023}. In this direction, our dataset representation techniques offer a first step toward data anonymization, enabling AI to process datasets without exposing the actual data. Future efforts to make AI more intuitive and reliable will further enhance the widespread adoption of modern AI data tools.

\textbf{Limitations \& Future Research.}
First, our future work focuses on enhancing \tool to better serve both general and specialized audiences. While basic chart types effectively communicate data to a general audience, more complex scientific use cases often require advanced and domain-specific charts that can encode multiple data dimensions. To address this, we plan to enable more personalized usage by allowing users to select advanced chart types tailored to complex, domain-specific analyses \cite{dibia2018data2vis}. Second, while \tool efficiently processes datasets with numerous rows, it encounters limitations when the number of columns increases, as the dataset representation must incorporate information about each individual column. To overcome this, we propose splitting datasets into smaller, more manageable tables automatically while preserving connections between them.
Lastly, we recognize that the general knowledge from LLMs may not be sufficiently in-depth to generate fact sheet for specialized scientific dataset. To ensure more accurate and insightful outputs, we recommend integrating Retrieval-Augmented Generation (RAG) techniques. This approach will enable \tool to leverage domain-specific knowledge bases effectively \cite{dibia2018data2vis}.

\section{Conclusion}

In this paper, we introduced \tool, an innovative system for automated fact sheet generation. This system is designed to create fact sheets from natural language requests and datasets. \tool generates fact ideas based on our proposed taxonomy and produces visualizations for each fact, which are then assembled into a comprehensive fact sheet. Additionally, we offer an AI-enhanced platform for users to personalize these fact sheets. User evaluations indicate a preference for \tool-generated fact sheets in terms of content detail and visual appeal. Further studies on our customization module have yielded positive usability results. Future efforts will concentrate on improving \tool's personalization capabilities and infusing it with domain-specific knowledge to cater for more advanced use cases.

\bibliographystyle{abbrv-doi-hyperref}

\bibliography{0main}

\appendix 

\newpage

\end{document}